\documentclass[letterpaper,twocolumn,10pt]{article}
\PassOptionsToPackage{hyphens}{url}
\usepackage{usenix-2020-09}

\usepackage[l2tabu,orthodox]{nag}
\usepackage{todonotes}
\usepackage{fancyhdr}
\usepackage{booktabs,caption}
\usepackage{threeparttable}
\usepackage{longtable}
\usepackage{array}
\usepackage{multirow}
\usepackage{lscape}
\usepackage{siunitx}

\begin{document}

\title{\centering Security Patchworking in Lebanon: \\Infrastructuring Across Failing Infrastructures}

		\author{}
	{
		\author{
			{\rm Jessica McClearn}\\
			Royal Holloway, University of London\\
			jessica.mcclearn.2021@live.rhul.ac.uk
			\and
			{\rm Rikke Bjerg Jensen}\\
			Royal Holloway, University of London\\
			rikke.jensen@rhul.ac.uk
			\and
			{\rm Reem Talhouk}\\
			Northumbria University\\
			reem.talhouk@northumbria.ac.uk
		}
		
	}

\maketitle

\begin{abstract}
In this paper we bring to light the infrastructuring work carried out by people in Lebanon to establish and maintain everyday security in response to multiple simultaneously failing infrastructures. We do so through interviews with 13 participants from 12 digital and human rights organisations and two weeks of ethnographically informed fieldwork in Beirut, Lebanon, in July 2022. Through our analysis we develop the notion of \emph{security patchworking} that makes visible the infrastructuring work necessitated to secure basic needs such as electricity provision, identity authentication and financial resources. Such practices are rooted in differing mechanisms of protection that often result in new forms of insecurity. We discuss the implications for CSCW and HCI researchers and point to security patchworking as a lens to be used when designing technologies to support infrastructuring, while advocating for collaborative work across CSCW and security research.
\end{abstract}

\section{Introduction}\label{sec:introduction}
Crumbling infrastructures, economic collapse, protests and political corruption are experienced on a daily basis by those living in Lebanon~\cite{AMAA2022,bahout2014,BouElA2016,abi2018,verdeil2018}. Stories of the impact of the economic crisis on the Lebanese population have been extensively reported in international media, e.g.~\cite{press:lebanesebanks:reuters}, with the spiralling economic collapse in the country labelled a ``ponzi scheme''~\cite{worldbank2022} and a ``deliberate depression''~\cite{worldbank2020} by the World Bank. A United Nations Special Rapporteur visit to Lebanon in November 2021 concluded that the political and financial leaders ``are responsible for forcing most of the country's population into poverty''~\cite{OHCHR2022} with 80\% of the Lebanese population having no access to basic rights such as healthcare~\cite{HRW2021b}. Research on these experiences highlights the challenges of negotiating an everyday existence in the face of failing infrastructures, while pointing to practices of resistance adopted by people both in and outside Lebanon~\cite{CSCW:ArmTalVla22,verdeil2018}; such as the 2015 and 2019 protests against the environmental and economic situation~\cite{Amnesty2019}. 

In this work we consider the impact of such failing infrastructures -- and specifically the infrastructuring taking place in relation to them -- on the everyday security practices among Lebanese populations. We explore the society-wide infrastructural failures in Lebanon from the perspective of how, through infrastructuring, people in Lebanon establish and maintain security in their everyday lives. In doing so, we bring to light the interplay between infrastructring and security that people in Lebanon experience as they navigate and negotiate multiple failing infrastructures. We draw on research that discusses infrastructuring practices when infrastructures and systems fracture and fail. Here, scholarship has referred to the infrastructuring done by, for example, patients in response to breakdowns in healthcare provision~\cite{CHI:GuiChe19,CHI:CVGMBBHP2019}, as part of community-based mentoring~\cite{CHI:DLILBRTWW22} or among solidarity movements~\cite{CHI:VCWO18}. In this body of work, the concept of infrastructuring is used to describe and analyse sometimes well-established practices developed to overcome particular institutional and systemic failures that are specific to the discrete contexts where infrastructures are lacking. Extending this work, we build on existing literature on \emph{seams and patches}, e.g.~\cite{vertesi2014seamful,CHI:SinJac17}, that encompasses the practices individuals, groups and communities enact as multiple infrastructures layer, connect and/or disconnect. Through the research presented in our paper, we bring these bodies of work into conversation with everyday security -- an understanding of security that is rooted in the ways in which daily routine activities provide continuity and, by extension, security. 
  
\paragraph{Contributions.}
Our work is ethnographically informed and grounded in fieldwork conducted over two weeks in Beirut, the capital city of Lebanon, in July 2022. We conducted 12 interviews with 13 people from digital and human rights organisations as well as activists who drew on their own personal experiences and their work with socially, politically and economically marginalised populations across Lebanon. Our research shows how the failures of multiple infrastructures -- electricity provision, public spaces, identification and authentication documents, access to money and financial resources as well as reliable Internet access -- necessitated infrastructuring to establish everyday security. We found that such routine practices simultaneously restored everyday security while leading to new forms of insecurity. This included, for example, additional and new economic concerns by using the services of unregulated `generator cartels' for electricity provision, relying on informal money exchange persons and platforms and the use of intermediaries to overcome limited Internet access. Accordingly, we propose \emph{security patchworking} as a lens that layers understandings of \emph{infrastructuring} with that of \emph{everyday security} and, in doing so, enables an analysis of infrastructuring as a form of everyday security in the Lebanese context and makes visible their intersections.

The research makes several contributions to CSCW scholarship. First, rooted in everyday security~\cite{nyman2021everyday,crawford2016mapping}, our work shows how fragmented systems, the simultaneous degradation of several State-level infrastructures and acts of non-governance in Lebanon led to unfulfilled everyday security needs that necessitated continuous infrastructuring by those living in Lebanon. Second, our work brings to the fore the complexities of infrastructuring performed to simultaneously secure access and services as well as provide everyday security. Third, our work presents the value of security patchworking as a lens that makes visible forms of security and insecurity as they pertain to infrastructuring work, the interplay between ground-up and top-down security infrastructuring and the ways in which infrastructuring for identity overlaps with other infrastructuring processes such as collective (online/offline) spaces, for example. Fourth, we discuss implications and potential directions of security patchworking for future CSCW work.
 
\section{The Lebanese Context}\label{sec:context}
We first outline the historical context of Lebanon to ground our work before highlighting the ongoing infrastructural failures taking place in the country. We present this context from the perspective of the participants' and our own understanding of how place and histories shape the situatedness of individual and collective interpretations~\cite{Haraway20}. 

In 1943, a newly elected Lebanese government rejected the French mandate and declared independence~\cite{Delatolla2021}. However, within the Euro-centric frameworks of state building and development, tensions in Lebanon arose and culminated in the outbreak of the Lebanese Civil War in 1975-1989, which was fought along religous sectarian lines. The Lebanese Civil War was also tied to wider geopolitical tensions in the region and saw the involvement of Israeli, Arab League and Syrian forces as well as Palestinian refugees displaced to Lebanon due to the Israeli-Palestinian conflict. The civil war ended with the signing of the Al-Taif Agreement in Saudi Arabia~\cite{NagClan2019}, which was supported by regional and Western powers. To this day, the Al-Taif Agreement is in force, with the sectarian warlords of the civil war forming the political elite that governs the country and has maintained societal sectarian divides~\cite{Salibi1990,Salloukh2019}.

Despite numerous loans and aid provided to the State of Lebanon for post-war infrastructural repair, the repairs have yet to materialise due to ongoing corruption~\cite{verdeil2018}. Investments in the electricity sector have been criticised for favouring government contracts along sectarian lines, with limited financial oversight. Power outages in Lebanon have become normalised~\cite{abi2021}, with households, hospitals and businesses having to cope with often only an hour of inconsistent State provided electricity a day~\cite{press:Lebanon:reuters}. Further, in 2020, ammonium nitrate stored in the Port of Beirut exploded, becoming one of the largest non-nuclear explosions in history~\cite{HRW2021}. The blast killed 218 and injured 7000 people, while destroying much of the city. A Human Rights Watch investigation~\cite{HRW2021} identified that corruption in the oversight of the publicly managed port allowed for the explosive substance to be unsafely stored under the cover of the opacity provided by the port's management structure. 

Corruption has also contributed to the failure of economic and financial infrastructures. The current spiralling economic collapse in Lebanon has been labelled a ``ponzi scheme''~\cite{worldbank2022} in light of the Central Bank of Lebanon providing loans to the government hedged by the US Dollar (USD) in the bank accounts of depositors in commercial banks. It is important to note that after the civil war, the Lebanese Lira (LL) was pegged to the USD by the Central Bank at an exchange rate of 1,500 LL to 1 USD with both currencies being used interchangeably within the country. With the government's inability and refusal to pay back loans to the Central Bank in 2019 people in Lebanon were notified that they would no longer be able to withdraw their savings from their USD accounts. Throughout the course of this research, the tensions between banks and depositors increased with arms being taken up in protest to withdraw funds that belong to the Lebanese population~\cite{press:Lebanon:reuters}. This mismanagement of the economy has led to hyperinflation and the LL losing 90 percent of its value in the time period 2019-2021~\cite{HRW2021b}. 

On 17 October 2019, people in Lebanon took to the streets to protest against the economic situation and corrupt mismanagement of the country. Protesters called for the political elite to resign under the slogans \emph{Kilon yaani kilon} (``All of them means all of them'') and \emph{Thawra} (``Revolution''). Protesters were met with beatings, teargas, rubber bullets and live ammunition by the Lebanese military and internal security forces. The same forces have also been reported to crack down on freedom of expression that was previously celebrated in the country~\cite{Amnesty2019}. The revolution is, arguably, still ongoing with protests happening in front of residences of politicians, commercial banks and government buildings, and it has become a nurturing ground for emerging new political imaginaries and parties. 

\section{Related Work}\label{sec:related-work}
We ground our findings in related CSCW, HCI and security work. Specifically, in Section~\ref{sec:rw-infrastructuring} we provide a synthesis of literature that highlights understandings of infrastructuring and the analytical vocabulary of \emph{seams and patches} as it has been used in relation to socio-technical infrastructures. Second, in Section~\ref{sec:rw-security-patchwork} we bring infrastructuring practices into conversation with everyday security.

\subsection{Fractured Infrastructures and Infrastructuring}\label{sec:rw-infrastructuring}
Scholars have explored the impacts of broken or fragmented socio-technological infrastructures on different user groups and populations. In so doing, they have highlighted the failures of the infrastructures themselves, while bringing to light the social and collaborative practices that surface when they fail -- often referred to as `infrastructuring work'. Infrastructuring terminology pertains to subtly different yet multiple conceptions; from infrastructuring being the tinkering of technologies by end-users~\cite{maclean1990user}, to the tailoring of technologies by practitioners~\cite{henderson1995there}, to infrastructuring being an ongoing process of ground-up (re)design~\cite{bjorgvinsson2010participatory} as a way to make platforms more accessible~\cite{simpson2023hey}. Literature has also looked at how infrastructures are built and maintained through collaborative, participatory design~\cite{prost2019infrastructuring,crivellaro2019infrastructuring,bodker2017tying,crabu2018bottom}, also how ``re-infrastructuring'' may occur to leverage existing infrastructures through introducing new technological capabilities~\cite{grisot2017re}. In this collection of work, infrastructuring is an assemblage of people, objects and technologies that come together to fulfil certain tasks~\cite{hussain2020infrastructuring}. Lee, Dourish and Mark~\cite{lee2006human} further expand on such definitions by highlighting the importance of human infrastructures in bringing together known and unknown people through formal and informal ties for the formation of routine engagements in order for work to be accomplished.

\subsubsection{Infrastructural Breakdown} 
When infrastructures break down their inner workings become visible~\cite{ruhleder1996steps} as their failure prompts people to engage with the assemblages and processes that constitute such infrastructures~\cite{ruhleder1996steps} and retrofit them over time~\cite{howe2016paradoxical}; this is often referred to as ``extrinsically motivated practice innovation'', where it becomes impossible to maintain existing practices and thus necessitating new and innovative approaches~\cite{ludwig2018designing}. For example, this might be seen when changes in pricing of a service or a technology make it unaffordable, thus requiring the use of new services or devices~\cite{ludwig2018designing}. Some authors have highlighted the ``point of infrastructuring''~\cite{pipek2009infrastructuring,ludwig2018designing}, referring to when users or practitioners of technological infrastructures realise that current usage needs to be reassessed, thus provoking ``in situ design activities''~\cite{pipek1999groupware}. More specific to CSCW and HCI, research has highlighted the ways through which infrastructuring takes place in socio-technical infrastructures in order to instigate and enact change~\cite{peer2022work}. Collectively, this literature hints at the ambiguity of infrastructure in practice as is also articulated in~\cite{mikalsen2018infrastructuring}. Our work sits within what is referred to as ``actual infrastructure breakdown'', where socio-technical systems become inoperable, as also highlighted by the authors of~\cite{joshi2021flaky}. Technologies have been used to navigate and overcome physical infrastructural disruptions during times of war and conflict, providing new routines and patterns of action for work and socialising~\cite{TOCHI:SemMar11}, safe spaces for recovery free from physical violence~\cite{SIGCHI:AlAMarSem10} and supporting people in collaborating with one another~\cite{SIGCHI:MarAlASem09}.

\subsubsection{Assemblages and Processes}\label{sec:rw-assemblages-processes} Karasti and Syrjanen~\cite{karasti2004artful} posit that by viewing infrastructures as assemblages and processes, the act of infrastructuring becomes that of the artful integration, co-ordination and meaning-making of people, objects and processes. In coping with infrastructural breakdowns, infrastructuring can be understood as ``maintenance work'', which not only relates to the maintenance of infrastructures but also to the anticipation and restoration of such infrastructures through socio-cultural practices, as shown in research on front-line health workers in South India~\cite{HCI:VBASMHG21}. The works of~\cite{CHI:GuiChe19} and~\cite{CHI:CVGMBBHP2019} further show how patients and caregivers engage in infrastructuring work to overcome barriers and breakdowns in institutional healthcare provision, while in~\cite{CHI:DLILBRTWW22} the authors present the notion of infrastructuring work within a community-based mentorship model building on social support, flexibility and trust.

More specific to infrastructuring within civic- and public-making endeavours, the authors of~\cite{peer2022work} highlight infrastructutring as a process of navigating the politics underpinning the work being done~\cite{korn2019infrastructuring}. Additionally, in resource-constrained settings, such as that presented in our paper, infrastructuring as a bricolage has been put forward~\cite{buscher2001landscapes}. This assumes that infrastructuring and those engaging in this process (bricoleurs~\cite{peer2022work}) will make use of what is socially, materially and culturally available to them. The authors of~\cite{CHI:VCWO18} further build on the notion of bricolage through their conceptualisation of `guerilla infrastructuring'. This refers to the creative leveraging and negotiating of resources and synergies from the periphery of centres of power through opportunistic, responsive, adaptable and decentralised tactics and strategies of making and re-making. Similarly, Holten M\o{}ller, Fitzpatrick and Le Dantec~\cite{holten2019assembling} highlight how people bring supplementary materials to the infrastructuring process as a form of ``private resourcing'' to complement infrastructures that build on existing power imbalances. 

Furthermore, others have explored different forms of solidarity infrastructuring that, while tangential to the specific focus of our paper, illustrate how collaborative practices enable and facilitate uneven infrastructures. For example, work has been done on how Rohingya refugees in Bangladesh use art-based practices to infrastructure hope~\cite{ICTD:HSSJA20} and on experiences of solidarity through digital and non-digital infrastructuring among sex workers~\cite{DIS:SMWCM20}. Additionally, the infrastructuring in and around failing State infrastructures has been theorised to contribute to new forms of citizenship that are decoupled from the State, as infrastructuring processes establish ownership and membership within the wider socio-technical system~\cite{joshi2021flaky}. Infrastructuring has also been described as acts of dismantling centralised infrastructures from below and, in doing so, creating inverse infrastructures that reveal power differentials, while enabling self-organised and de-centralised control amongst stakeholders~\cite{egyedi2012inverse}. Further, as posited in~\cite{howe2016paradoxical} and exemplified by~\cite{singh2021seeing}, thinking with and through infrastructures, as we do in this paper, allows for infrastructural failure to be used as an analytical lens~\cite{MiyRil05} to examine inclusion and power.  

Collectively, the above scholarship highlights the ways through which infrastructuring enables communities and individuals to regain control over services~\cite{joshi2021flaky,peer2022work}, their personal experiences~\cite{semaan2019routine}, their economies~\cite{CHI:VCWO18} and futures~\cite{holten2019assembling,joshi2021flaky,CHI:VCWO18,peer2022work,ICTD:HSSJA20}. Additionally, in the case of~\cite{rifat2022}, the research shows how infrastructuring work is done to create and disseminate Islamic content in the absence of digital faith infrastructures. What the above studies also highlight, among other things, is how fragmented infrastructures and systems are made to work for people through distinct social and collaborative practices pertaining to their specific contexts. In~\cite{HCI:JGVCPMJEADBH19} the authors note, in their study on the infrastructuring of technology repair in low-income areas in the Philippines, how informal practices grounded in trust relations emerge.

\subsubsection{Seams and Patches} Vertesi's~\cite{vertesi2014seamful} conceptualisation of \emph{seams} as an analytical vocabulary has also been used when articulating infrastructuring. It is used to provide an understanding of how actors ``artfully align'' infrastructural commitments in an effort to achieve certain goals by bringing together objects and processes. This is done to work within gaps that arise when multiple and heterogenous infrastructures partially overlap, connect and/or disconnect~\cite{vertesi2014seamful}. Within this vocabulary, seams are the visible edges, endings and gaps between and across multiple systems. In line with Vertesi's notion of seams, associated terms such as \emph{seamful} and \emph{seamless} have been used to articulate infrastructuring work. Here, seamful refers to how assemblages are visible and felt by users of systems especially due to their unevenness, while seamless refers to the invisible, in-the-background assemblages that stitch across systems. Singh and Jackson~\cite{CHI:SinJac17} point to how the use of seams as an analytical construct makes visible the challenges that arise when layering a technology over a pre-existing system. Additionally, they note how not everyone is equally able to navigate the seams of such layered infrastructures, which results in people getting `stuck' in seamful spaces. It is with this understanding that infrastructural seams determine inclusion as a process of navigation, rather than a binary existence. The analytical use of seams further enables scholars to identify the different stakeholders that stitch across multiple technologies, while allowing them to unpack ``one seam after another''~\cite{CHI:ChaKum18}.  

Vertesi~\cite{vertesi2014seamful} also borrows from `software patching' to highlight how actors produce fleeting moments of alignment within seams that do not necessarily come together as a stable whole but can be viewed as a patchwork that loosely sews across seams. Further, Dailey and Starbird~\cite{dailey2017social} note how during crises, social media users become ``seamsters'' by assembling different social media platforms in an effort to meet their information needs. Here, the configuration of multiple social media platforms and their use constitute a patchwork of patchworks with each patch serving a different informational function~\cite{dailey2017social}. Adopting a top-down view, the authors of~\cite{veeraraghavan2021cat} refer to patching as an iterative and political process of ``adapting information infrastructures to deal with problems of implementation''. Here, the authors highlight how government bureaucrats use technologies to reconfigure bureaucratic procedures in order to meet particular political goals and to respond to policy, implementation knowledge and resistance from below. Conversely, taking a ground-up position, the authors of~\cite{fox2023patchwork} use the concept of patchwork to refer to the human labour -- such as calibration, troubleshooting and repair -- undertaken to fill the gap between what AI technologies are supposed to do and what they actually do. Of particular relevance to our study is work that brings together conceptualisations of infrastructuring and seams when examining State infrastructures. Here, Joshi, Bardzell and Bardzell~\cite{joshi2021flaky} show how the lack of seamless functioning of water infrastructures in Pune, India, coupled with the economic and political inequalities in water service provision, contribute to the lack of legitimacy of the system. This makes the failing infrastructure itself, along with its assemblages and processes that reside within its seamful spaces, visible for scrutiny. However, their research also found that the ``accretions of cultural practices, jugaads and appropriations so delicately interwoven through everyday mundane actions''~\cite[p.20]{joshi2021flaky} to make the water infrastructure function at a sufficient level, re-invisibilises it. This, they refer to as ``infrastructure (re-)gained''.

\subsection{Everyday Security}\label{sec:rw-security-patchwork}
The everyday has become a category of analysis in security studies in recent years, with scholars highlighting how daily routine activities provide continuity and, by extension, security. A focus on the everyday moves our engagement with and understanding of security away from high politics and the `spectacular' to more mundane and intimate relations, interactions and practices. As noted by Nyman~\cite{nyman2021everyday}, security shapes everyday life and is an integral part of people's lifeworlds; it shapes the spaces they inhabit, the activities they undertake and the experiences they have. In~\cite{crawford2016mapping}, Crawford and Hutchinson explore what they refer to as ``security experiences'', foregrounding practices through which people \emph{do} and \emph{experience} security. In so doing, they exemplify how such experiences are established through and find expression in ``quotidian aspects of social life''. Building on Crawford and Hutchinson~\cite{crawford2016mapping}, and synthesising existing scholarship on everyday security across disciplinary lines, Nyman~\cite{nyman2021everyday} highlights that ``the everyday life of security is multifaceted and exists in mundane spaces, routine practices, and affective/lived experiences''. The experiences of security (and insecurity) as part of people's daily lives thus shape how it is taken up, resisted and augmented by individuals and groups in everyday settings. Such an approach to and understanding of security focuses attention on the micro and proximate interactions and relations that make up the everyday, rather than the formal, official processes of the State~\cite{vaughan2016vernacular}. As this body of scholarship suggests, everyday security broadens the security conversation to respond to the enmeshed nature of technology and society `on the ground'. Everyday security thus brings into conversation different forms of security. It allows for an analysis that considers a broad definition of security, while supporting an understanding of identity in security terms. Here, identity is a dialogical practice through which people can establish a sense of ontological security~\cite{cunliffe2001,mcsweeney1999},\footnote{The concept of \emph{ontological security} was developed by sociologist Anthony Giddens~\cite{giddens1991}.} which Steele~\cite{steele2005} refers to as ``security as being'' and Mitzen~\cite{mitzen2006} as ``security of the self''.

Recent work published in CSCW and HCI has employed the notion of everyday security in, for example, the contexts of isolated communities at the margins of society~\cite{CHI:JCWL20} and refugees (re)settling in and accessing a new country~\cite{CHI:ColJen19,CHI:ColJenTal18}. Thus, CSCW and HCI scholars have sought to understand security as grounded in social, and often creative or resilient, socio-technical practices. In particular, some situated research has explicitly foregrounded social contexts, structures and relations to explore how these are used to respond to technological security issues. For example, in~\cite{CSCW:KKPW18} the authors used the notion of care as an analytical framework to emphasise the often invisible and feminised practices that underpin IT security; what they refer to as ``tinkering''. Further, in~\cite{USENIX:GCKRC21} the authors explored how video chats within close-knit social groups could be used as an authentication method, highlighting the impact of the specific social context and relational ties on security. In~\cite{HFCS:VBDVTMO12}, the socio-materiality in relation to security were foregrounded in the context of financial security, while the authors of~\cite{CHI:OITD19} showed how feelings of insecurity shape online practices. Additionally, in their influential work, Dourish and Anderson~\cite{HCI:DouAnd06} situated security and privacy questions within social and cultural contexts. 

In the studies outlined above, everyday security is not understood as an attribute inherent to a specific technology or system; rather, everyday security, the perceptions thereof and the associated practices, is understood in relation to its underlying social dynamics, processes and (infra)structures -- as well as the routine work performed by individuals and groups to negotiate and maintain security in their everyday lives. Everyday security finds direct support in Semaan's~\cite{semaan2019routine} notion of ``routine infrastructuring'', where marginalised groups work to regain control of elements of their daily lives through continuous infrastructuring. Through routine infrastructuring, the building of resilience becomes a routine activity through which individuals aim to (re)gain security in their everyday. In security terms, everyday security, as it is rooted in the mundane and proximate, can be understood as a parallel to infrastructuring within CSCW and HCI scholarship. Further, everyday security itself can be seen as a form of infrastructuring or `repair work' as argued by the authors of~\cite{korn2017friction}. 

\section{Methodology}\label{sec:methodology}
The research combined semi-structured interviews with ethnographically informed fieldwork\footnote{We use \emph{ethnographically informed} to highlight the short-term nature of the fieldwork.} in an attempt to overcome the inherent bias in studies that rely solely on interviews as participants self-select to take part~\cite{USENIX:ABJM21}. The research was further contextually formulated and informed in collaboration with a Lebanese researcher who has conducted extensive fieldwork in Lebanon. Fieldwork in Beirut was carried out by one researcher over a two-week period in July 2022, where the researcher stayed in a studio apartment in Mar Mikhael, Beirut -- a diverse part of the city that was one of the most impacted areas by the 2020 port explosion. The researcher participated in several events while in Beirut; for example, as an audience member at a poetry event, at an open mic night as well as on a walking tour of the city. It was through such interactions that the subtleties of the different infrastructuring practices were observed \emph{in situ}.

As part of the fieldwork, 12 semi-structured interviews were conducted with 13 digital and/or human rights workers; nine in person and three online (see Table~\ref{tab:participants}). The in-person interviews were conducted in locations where participants felt most comfortable, for example, in local coffee shops or in office settings. All interviews were conducted in English. Despite Arabic being the national language of Lebanon, English and/or French languages are taught in Lebanese schools -- including public schools -- and are widely used in Lebanese workplaces. While in Beirut, the researcher noticed how conversations were often conducted in all three languages with individuals seamlessly switching between languages mid-sentence. An interview guide (see Appendix~\ref{app:topic-guide}) was used and topics included the exploration of privacy and security in relation to online access in Lebanon, the availability and accessibility of public spaces as well as access to digital spaces. Despite the construction of an interview guide, the interviews were purposefully kept broad so as not to `force a security angle', which was well suited for our research especially as it became evident that participants were consistently speaking to issues of security using the language of infrastructures and infrastructural failure. 

\begin{table*}\centering
  \caption{Overview of interviewed participants as part of the fieldwork in Beirut.}
  \label{tab:participants}
  \begin{tabular}{cclc}
    \toprule
    ID & Organisation type & Interview location & Interview length \\
    \midrule
    P0 & Human rights-focused foundation& in office& 74 minutes \\
    P1&Human rights-focused foundation& in office& 74 minutes \\
    P2& Education and social change initiative& in coffee shop& 39 minutes\\
    P3& Political party worker& in office& 53 minutes\\
    P4& Digital rights and design researcher& online & 34 minutes\\
    P5& Internet governance and technical expert& in coffee shop& 41 minutes\\
    P6& Democracy and digital rights organisation& in office& 71 minutes\\
    P7& Digital rights organisation& in coffee shop& 46 minutes\\
    P8& Democracy and elections organisation& in office& 93 minutes\\
    P9& Independent media organisation & online& 31 minutes\\
    P10& Diaspora and human rights& in coffee shop& 69 minutes\\
    P11& Digital rights organisation& online& 24 minutes\\
    P12& Digitalisation organisation& in office& 47 minutes\\
    \bottomrule
  \end{tabular}
  \begin{tablenotes}
  \small
  \item \small \emph{Note:} We do not refer to participant IDs throughout the paper to avoid inferring identification. For a similar reason, we only provide high-level descriptions of the organisation types. P0 and P1 were from the same organisation and took part in the same interview.
  \end{tablenotes}
\end{table*}

\paragraph{Participants and recruitment.}
There were no specific selection criteria for interview participants apart from them having an interest in and knowledge of digital rights, human rights and/or related socio-political issues within Lebanon. Three interviews were confirmed before arriving in Beirut -- despite contacting 37 organisations and individuals -- while the remaining nine interviews were established during the fieldwork. As well as interview participants, this research would not have been possible without the countless interactions with those encountered in Beirut; conversations with taxi drivers, baristas and those met in different contexts. In some ways, this research follows Nader's~\cite{nader1972} approach to ``study-up'' by engaging with participants who hold some power in the context of the specific research topic. Participants who were engaged through formal interviews, while not decision makers in official policy terms, held seats at the table in terms of debates relating to (digital) access and rights, including representatives from Non-Governmental Organisations (NGOs), charities and political parties. Interview participants were able to draw on their own personal experiences of living in Lebanon as well as their knowledge and experiences of working with marginalised populations in Lebanon. As such, while they may personally be in a better socio-economic position than those most impacted by infrastructural failures, through their work they were able to provide insights on the infrastructuring practices of the communities they support and work with. 

\paragraph{Ethical considerations.}
The research received full ethical approval from our institution's Research Ethics Committee. Interviews were audio-recorded, before being transcribed and anonymised, and subsequently destroyed. Observations, informal engagements, personal reflections and emotions were captured in the form of field notes, and incorporated in the data set underpinning this paper. The research adhered to data minimisation practices in terms of personal identifiable information about participants. All interview participants received a Participant Information Sheet prior to the interviews and all signed a consent form. The data is securely stored in line with our institutional policies. 

\subsection{Data Analysis}\label{sec:data-analysis}
Interview and field-note data was manually annotated and grouped into themes, followed by an inductive approach to look for subtleties that enabled us to obtain a more in-depth understanding of the data. The analytical approach was rooted in Braun and Clarke's~\cite{braun2019reflecting} approach to reflexive thematic analysis, which recognises that the themes are shaped by one's own positions and perspectives and thus rest on a series of decisions made in the analytical process and not solely on the count prominence of codes and themes. The analytical process also involved collaborative analysis of the anonymised data among the co-authors, through long analytical discussions, where the constructed themes were challenged and brought into conversation with each other as well as broader scholarship. Collaborative analysis also continued throughout the write-up stage which is considered to be one of the analytical steps of thematic analysis~\cite{braun2006}. As a result of this collaborative analytical approach, themes were deepened, revised and refined. For example, data pertaining to the failure of the electricity infrastructures in Lebanon, the workarounds people employ such as relying on electricity from generator cartels and solar energy, and the exclusions and insecurities such workarounds entail were coded as `failing electricity infrastructures', `workarounds employed to secure electricity' and `insecurities electricity workarounds entail', respectively. They were then grouped together to construct the theme `electricity infrastructuring' which in writing the paper was further grouped under `Failed Infrastructures and Material (In)Security' (see Appendix~\ref{app:coding-table}). Importantly, however, relaying the interactions that occurred through the fieldwork can only merely attempt to convey the experiences themselves. This limitation is also reflected in Plow's~\cite{plows2018} discussion on ``messy ethnographies'', highlighting how presented findings often appear as a ``sanitized account'' of what happens during fieldwork. Additionally, we acknowledge that while we analyse and recount the researcher's experiences and interactions in Beirut as they navigated dilapidated infrastructures, their experiences are not fully comparable to those of people living in Lebanon.  

\paragraph{Researcher positionalities.} Our multiple researcher positionalities collectively shaped the research approach and analysis. They constitute what others have referred to as a ``patchwork of positionalities''~\cite[p.38]{PDC:CTBBKBGB2022}. In our research, one researcher carried out the fieldwork and shared the positioning of Nicola Wendt in~\cite[p.3]{wendt2020}. Wendt commented on her position as a ``white, female, junior researcher'' who had no prior links to the groups under study and the need to ``place the everyday experiences and concerns of the participants at the heart of the research design to avoid any kind of Euro-centrism''. As part of the individual positionality we carry, the researcher who spent time in Beirut, was born in Northern Ireland. Therefore, she identified with participants recounting political frustrations and the lived experiences that shape lives in a country historicised through and with conflict. The researcher team also included a Lebanese researcher whose interpretations of the data are intimately tied to her lived experiences within the Lebanese context, and researchers with knowledge of participatory design and mundane security practices in spaces of marginalisation and higher-risk.

\section{Findings}\label{sec:findings}
Our research points to the failings and dilapidation of multiple, intertwining and large-scale infrastructures that were foundational to the everyday lives of participants, such as utility supply and economic infrastructures. This was captured by one participant who summarised: ``Infrastructure, what infrastructure?''. This dilapidation was experienced on a daily basis and drove extensive infrastructuring work by people in Lebanon, at both an individual and collective level. While not directly referred to in security terms by participants, the ways in which failing infrastructures were talked about highlights the deep-rooted everyday insecurity experienced in Lebanon. In this section, we outline our findings in this context. We do so in three subsections, with each of which foregrounding distinct aspects of the intersection of infrastructuring and security. First, Section~\ref{sec:infrastructure} establishes the foundations of the infrastructuring practices that people in Lebanon undertook for their own material security. Second, Section~\ref{sec:spatial-security} articulates how infrastructuring was undertaken to establish digital access and security. Third, in Section~\ref{sec:identity-security} we report on findings relating to how infrastructuring was done to secure identity in the Lebanese context.
 
\subsection{Failed Infrastructures and Material (In)Security}\label{sec:infrastructure}
A key finding from our analysis relates to how people in Lebanon engaged in infrastructuring to ensure their own material security that was not provided by the State, i.e. their essential and basic needs. This also meant that questions related to (digital) privacy were less pressing for participants in our study as they emphasised how the society-wide crisis had led to Lebanese populations focusing on survival: ``If you look at the news people are fighting to get bread, to have electricity and gas'' and ``of course people care about their privacy but it is not a priority now.'' In a similar vein, privacy was often referred to as a secondary thought or a ``luxury''. As one participant, who worked on digital rights and privacy matters, noted: ``Priorities right now for most people are basic necessities rather than maybe talking about privacy laws.'' Instead, practices of daily survival underpinned conversations and encounters around security and participants gave several examples of how people in Lebanon had developed different, ground-up approaches in attempts to overcome their experiences of material insecurity.

\subsubsection{Infrastructuring Ponzi Economics}\label{sec:infrastructuring-ponzi} The economic crisis in Lebanon was foregrounded in everyday encounters in Lebanon and directly linked to the insecurity experienced by those living in the country. As noted in Section~\ref{sec:context}, banking services in Lebanon shut down in October 2019 when the protests initially started and have yet to fully reopen.\footnote{This is correct as of the time of writing, July 2023.} This economic shutdown was felt in everyday interactions in Beirut as people engaged in different forms of infrastructuring to attain the Lebanese currency. This was experienced through the fieldwork as well. One of the first actions upon arriving in Beirut was to exchange cash, as it was not possible for the researcher to use an ATM due to restrictions on cash withdrawals. The insecurity created by not being able to access financial resources was `resolved' by contacting a local exchange person via WhatsApp to deliver the requested cash. The WhatsApp number for this person was provided by the Lebanese co-researcher who obtained the contact details based on the recommendations of friends living in Beirut. After a short period of time, the person arrived on a moped with Lebanese Lira stashed under the seat. They then exchanged the researcher's USD using the market rate.\footnote{During the time of the fieldwork, the official rate of exchange was 1 USD to 1500 LL yet the market rate was 1 USD to 32,100 LL. This has since increased three-fold with 1 USD equating to 93,200 LL in June 2023.} This experience exemplifies the regular practices undertaken by those living in Lebanon, where access to personal funds is restricted, with a maximum withdrawal allowance of approximately 400 USD per month. The research also exemplified how such currency exchanges and withdrawals had become routine practices for Lebanese people. Such practices bypassed the official USD to LL exchange rate set by the Lebanese government, in favour of the market rate which was shared online via multiple websites (e.g., lirarate.org). However, we observe how on such platforms exchange rates fluctuated daily if not hourly with little to no information provided -- and a lack of transparency on how the rate is being formulated and on the individuals, groups and/or companies that are sustaining these platforms.  

Infrastructuring ponzi economics was also tied to the wider electoral system in the country. Here, participants indicated how Lebanese citizens would often accept having their votes bought by political parties for financial gain. Indeed, participants suggested that ``vote buying'' was instrumental to election outcomes. One participant involved in local politics commented on how this had been exacerbated throughout the economic decline and how the purchasing of votes had become cheaper: ``There was more money in the country in 2018 than there is in 2022. The votes are cheaper [now]. I remember I was offered 133 USD [for my vote] during the last elections, this election the amount was between 30-50 USD.'' Vote selling thus became a form of infrastructuring economic security within the corrupt political system. 

\subsubsection{Electricity Infrastructuring}\label{sec:infrastructuring-electricity} The electricity sector in Lebanon is ``broken'', as noted by several participants, meaning that State-provided electricity could only be expected for approximately two hours per day.\footnote{There was no consensus among participants as to the number of hours they could expect to have government-provided access to electricity on a daily basis.} One consequence of electricity outages in Lebanon was explained to be ill health and the flickering electricity of, for example, fridges was also observed during the fieldwork. Participants indicated that due to the electricity outages there had been increased cases of food poisoning: ``I had food poisoning for like two months because there is no electricity in the country and our food was going off. So you know I was subsisting on a diet of toast and [canned] tuna for a very long time.'' The researcher conducting the fieldwork also struggled to find fresh fruits and vegetables in Beirut. Due to the frequent power outages, she would also often be grocery shopping in places with no electricity. This meant that often pen and paper would be used instead of a cash register, while mobile phones functioned as flashlights. 

Several participants noted that those with political power in the country would have full access to electricity as well as other essential services that ensured their material security. It was also clear from engagements with participants in Beirut that wider political and religious sectarian discourses and divisions dictated electricity provision in the country. As a result, many in Lebanon were reliant on electricity provided through generators where provision was increasingly controlled by what some participants referred to as the ``generator cartels''. In many ways, these cartels helped to bridge -- or \emph{infrastructure} -- the fragmented electricity system by providing services which the Lebanese authorities did not. Yet this had given space to an uncoordinated and unregulated market controlled by such cartels, which was explained to be costing people large sums of money and contributing to further financial insecurity. Participants also indicated that those with financial means had turned to solar energy for the provision of electricity: ``You see a lot of solar panels when you are walking around, on buildings and everything, you would not see this last year''. Additionally, one participant reported how their organisation was ``partnering with a solar energy company which is offering to the schools a 50\% discount and the second 50\% they need to pay it''. However, despite the infrasturcturing work undergone to secure electricity, our fieldwork highlighted that this was not a seamlessly successful solution to lessen the impact of material insecurity created by State-level infrastructural failures. Only those who could afford this form of infrastructuring by installing solar panels or keeping up with the rising prices of the ``generator cartels'' were in a position where they could benefit from these workarounds. Furthermore, one participant noted that these workarounds were often hard to navigate given that:  ``We do not have a platform that can bring these people [such as generator owners and solar panel providers] together and make these services together for critical mass''.

\subsection{Infrastructuring Digital Access and Security}\label{sec:spatial-security}
Beyond the cost of electricity provision, one participant commented that the cost of some phone and data packages had reached the monthly minimum wage in Lebanon: ``Many people, due to the economic crisis, will be disconnected''. The coverage was also said to be dropping with the Lebanese government considering ending 2G services and reducing the 3G network by 90\%. As expressed by participants and highlighted in recent reports~\cite{SMEX2022}, with no 2G service, approximately 300,000 people would be excluded from the mobile phone network. Particularly, participants stressed that this would largely impact already underprivileged communities who relied on the 2G network. The failing Internet infrastructure and its impact on marginalised communities had already resulted in a reliance on intermediaries for accessing essential digitalised services such as accessing aid and COVID-19 vaccinations: 

\begin{quote}
    ``The people who register for a vaccine or for aid are the ones who have access to Internet or a phone. What we [the organisation] had to do in this case was create a second layer of outreach through NGOs. We connected with NGOs and created a portal with them, here you have to go through intermediaries. In some other cases we went through Mukhtars, the locals.''\footnote{\emph{Mukhtar} refers to the administrative head of a village or neighbourhood in Lebanon, as well as in many Arab countries.} 
\end{quote}

\noindent Our analysis also revealed several examples of how people in Lebanon would often engage in multiple practices in an attempt to overcome the intersection of the failing electricity and Internet infrastructures. One participant explained how people would rely on coffee shops and neighbours to connect: ``If they do not have electricity, they put the charger in some public place to charge then they go to the neighbour to connect to WiFi''. This comment highlights how such practices had become routine activities for those without access to reliable electricity or WiFi. Such practices were grounded in the lack of economic means, while bringing to light how they shaped the everyday lives of many people in Lebanon. It further highlights the division between those with reliable Internet access and those without -- grounded in economic means -- exemplified here by the participant referring to ``they'', i.e. not themself.

The above examples show how infrastructuring work took different forms and existed on multiple levels in parallel (material, electricity and Internet provision), while it relied on routine practices and relational ties within neighbourhoods and local communities. It was evident from multiple encounters in Beirut that such practices were adopted as direct responses to infrastructural failures, while attempts at mitigating their impact happened at a local, community level. Such infrastructuring meant that people had to rely on each other to obtain digital access and security. Similar to how the lack of electricity provision was experienced throughout the fieldwork, so was limited and unreliable Internet availability which was witnessed in most everyday interactions in Beirut. For example, at an open mic night, it was observed that the public WiFi password was, rather humorously but tellingly, ``slowWIFI4all''. In some ways, this can be seen as an insignificant observation during a two-week field visit, yet, it speaks to the resilience with which everyday infrastructuring was undertaken by those living in Beirut. It further underpins Semaan's notion of ``routine infrastructuring''~\cite{semaan2019routine}.

\subsubsection{Infrastructuring Ownership of Online Spaces} State ownership of online spaces caused everyday security concerns among participants. Despite there being multiple Internet providers, these companies were said to be owned by a centralised government body. This led some participants to stress that they did ``not want [the] government anywhere near the Internet''. Further, there was a sliding scale of perceptions of security and freedom of speech among participants; from one participant saying ``we have, and we enjoy our freedom of expression but with many caveats'' to another noting ``I mean, if you say something wrong about the President you might end up in prison''. Others referred to State ``policing'' of online spaces. The perception of being monitored extended beyond State-related surveillance to include supporters of the political elite. Some participants were particularly concerned with their online presence and of certain groups within Lebanese society due to the atmosphere becoming increasingly ''violent''. Participants cited recent government-driven online movements against LGBTQ+ groups, Syrian refugees, women, and journalists. One participant stated that ``everyone is watching everyone, and it is not a safe environment''~\cite{USENIX:MccJenTal23}.

Participants further noted how collectives or `bubbles' were formed online as a way for people to secure their own online presence by, for example, following those who shared a similar world view: ``If you tend to follow a lot of independent media outlets, that is my bubble, but someone like a family member who lives in Baalbak but may be very conservative or pro (extreme political views) will follow the [anonymised political party]  channels, whether that is on Facebook, Twitter.'' For participants, following those with shared world views was seen as a form of taking ownership of online space.  

Several infrastructuring practices undertaken to counter State control were recounted by participants and were said to exist at multiple levels in parallel. For example, participants highlighted how the economic collapse had resulted in a series of independent voices emerging to counter State-governed narratives and to infrastructure online spaces and ponzi economics: 

\begin{quote}
    ``They were on Twitter and experts in finance, the economy and energy or whatever and it was a loose grouping called the NERDS, which stood for Nasty Economics Requires Drastic Solutions. So these guys were briefing the public, they would look at anything they could access in terms of information from the central bank and they would put out a series of tweets and say this is the state of our economy, this is what needs to happen and it was information we weren't getting from the media or our own government obviously so they played a really critical role.'' 
\end{quote}

\noindent As highlighted by participants, times of social and political pressures forced communities to find new means to share knowledge and experiences, which often resulted in different forms of online infrastructuring as exemplified here. This was illustrated by another participant who discussed multiple innovations which appeared in the education sector. One example brought into conversation everyday security related to access and ownership of online spaces: ``Because of the issue with the lack of Internet connection, they developed something called Lebanese Alternative Learning in a Box, which you put in the school without Internet and it becomes like a hotspot. It enables students to go into their [Lebanese Alternative Learning] websites and work within it.'' These examples highlight the ways in which people in Lebanon were infrastructuring to take ownership of online spaces either through taking up space in existing social media platforms or by creating new spaces for others to access.

\subsection{Securing Identity through Infrastructuring}\label{sec:identity-security}
In this section we bring into conversation the infrastructuring practices that participants reported in relation to securing a sense of identity (Section~\ref{sec:lebanese-identity}) as well as material manifestations of identity autentication such as passports (Section~\ref{sec:physical-identity}). These practices were directly linked to society-wide infrastructural failures, including that of public space, which required the establishment of multiple and overlapping routine activities to build and maintain everyday security. 

\subsubsection{Identity Infrastructuring}\label{sec:lebanese-identity}
The notion of identity is fundamentally linked to ontological security with infrastructuring becoming one way of trying to achieve ontological security through a negotiation of one's identity~\cite{semaan2019routine}. A desired sense of ontological security may thus also lead to particular forms of identity work. This includes, for example, the construction of an identity that holds some form of power or ensures a sense of belonging -- either as an individual or collective. For the participants in our study, notions of identity were particularly contested. ``We do not have a collective identity as Lebanese people'' was noted by one participant, while others commented being constantly conflicted in identity terms: ``We are all living an identity crisis in whether we belong to this country.'' The struggle for identity was rooted in the lack of essential services creating a sense of each sect and/or region having to fend for themselves, with one participant articulating how such traumatic experiences shaped notions of identity: ``The experience of being violated, constantly being violated is something that people live with and deeply internalise''. 

The ``continuous reflection about our identity'', as noted by another participant, was also reflected in people's everyday security responses and needs. One illustration to this point is how people responded during the 2022 elections. Participants anecdotally shared how, on occasions, securing identity was a form of gaining material security and took precedent over securing the political imaginaries of the revolution. One participant referenced a family member who voted for a traditional sectarian political party that is part of the political elite, instead of new political parties emerging from the revolution, despite ``cursing them throughout the year.'' The participant, paraphrasing this family member, described: ``When I vote for them I feel secured as I am voting for the State. They give me services. If I don't vote for them and I vote for people with an agenda I agree with, I might end up being starved.'' The participant explained that vouchers and aid were often provided in exchange for electoral political support. This example underlines how economic hardship influenced people's everyday security decisions which were intertwined with notions of identity and, by extension, their ontological security. 

However, the above example should also be understood in the wider context of identity, which is intertwined with space in the Lebanese context. Participants highlighted that their experienced identity crisis was reinforced by how space within Lebanon in general and Beirut in particular is divided along sectarian lines. Space in Lebanon has always been contested and, for years, physical demarcations of space have been used as a tool of social control in the country. During the Lebanese Civil War, public spaces and urban areas were divided along sectarian lines and the identity divides associated with them~\cite{Salibi1990}. These spatial divides have been maintained since then, with one participant capturing this: ``Urban planning is associated to identities and to where you belong.'' Furthermore, the lack of public spaces such as parks and libraries was noticeable during the fieldwork. The researcher observed that public spaces in Beirut were limited and that coffee shops were often used as gathering points for small groups of people while others inhabited public space when they sat outside their own stores. Additionally, several participants explained how the COVID-19 pandemic had been a reason cited by the Lebanese authorities to close the few existing public spaces in Beirut, and most of the few parks in the city remained locked after COVID-19 restrictions were lifted. Such public spaces had previously functioned as particular infrastructures for people to come together in the city, but had become another example of a failing infrastructure. When asked whether they felt that the lack of public spaces meant people inhabited online spaces more as a means of coming together, participants generally accepted this interpretation. One participant noted: ``That is a very good observation, so you are saying because there are not enough public spaces, people are ending up spending more time online.''

Participants narrated that this crisis of identity and space was temporarily resolved in ``a really beautiful, unprecedented moment despite the hardship'' during the 2019 revolution when economic hardship and frustrations over political corruption overrode sectarian divisions. Here, people re-took urban spaces as an infrastructure by inhabiting and taking ownership of physical, privatised, spaces in the city: ``We were all just going down [to those spaces], screaming for livelihood [\dots] calling for a State to fit us all.'' This highlights how occupying spaces together was a means of infrastructuring towards securing a sense of collective identity. This form of infrastructuring was mirrored by similar practices online. Online spaces often reflected and sometimes intensified the societal and sectarian challenges felt by participants and across Lebanese society, thus, creating greater polarisation: ``If you are sitting in the same space with different people physically you may realise you have more things in common than being in your online bubble.'' However, the use of online space -- or ``ecosystem'' as articulated by one participant -- was felt by multiple participants as an important support: ``The online space has been this circle of friends and we create our own reality.'' Through our analysis of the data, it became clear that online platforms often became spaces where younger Lebanese people would share their experiences and support each other through a collective grieving process, rooted in their experiences of the 2019 protests in particular. The movement to online spaces, partly due to a lack of publicly accessible collective infrastructures such as public parks and partly due to a population scattered by conflict, thus underpinned the practices undertaken to infrastructure towards a form collective identity. As noted by one participant: ``I became much more active on Twitter during these few years because I felt if I was able to share my pain with others and they were able to share it with me I would somehow feel a lot less alone.'' 

Lebanese diaspora communities also found ways of engaging in the infrastructuring work that underpinned the protests and sharing the grief which surrounded them: ``I would be at work, I had my two screens, laptop and two phones and all of them would have something on them, the news, Facebook live, two different accounts of Instagram live, everyone was stuck to their screens wanting to know what was going on [in Lebanon].'' These spaces, despite allowing a form of remote participation, also created a feeling of helplessness and guilt for some as they watched loved ones take part in the protests: ``I will never forget how I saw my friend got beaten up in front of me on the screen and I could see him getting kicked on the head.'' It was also clear from the encounters in Beirut with diaspora communities -- visiting the city at that time -- that they were constantly navigating their own understandings of place and identity through the online representations of trauma, violence and insecurity experienced by people ``left behind'' in Lebanon. For example, it was noted that video conferencing platforms such as Zoom were referred to as ``places to `sit together' '', sometimes for several hours without talking but to share the suffering experienced by friends and family in Lebanon and through such shared experiences secure a form of collective identity, albeit at a distance. 

\subsubsection{Securing Manifestations of Identity and Authentication}\label{sec:physical-identity}
Fieldwork and interview data also exemplified identity-specific infrastructural failings across Lebanon, with participants articulating different forms of infrastructuring in this context. For example, due to the fragmentation of government departments separate forms of ID were required for, e.g. a passport, the tax office, the police, vehicle registry, to mention a few. One participant explained how they had seven different identity documents, some of which were remnants of paper-based identity governance systems while others were digitised. Several participants suggested that the infrastructural failings and reliance on paper systems ultimately served corruption by the political elite, with one participant indicating that the paper-based systems further enabled corruption: ``If you [referring to the researcher] want to do any paperwork there is a system that is transparent and you can follow it up? Here, it probably disappears or you have to pay somebody [to successfully submit the paperwork].'' It was also noted that the economic collapse had meant that the government could no longer afford paper to print passports and/or pay for fuel to supply the printers with electricity. This was in addition to participants explaining that recently implemented digitised systems were slow, resulting in extensive delays, which meant that people were waiting for over a year for a new passport. One participant explained: ``People are not being issued passports, I mean it is the basic function of a State to provide you with identity papers.'' The multiple issues within the passport system highlight the dilapidation of the government functionality in Lebanon: ``Those who cannot leave are literally being held hostage by a State that is crumbling and an economy that is collapsing.'' This feeling of being trapped was also experienced through participant observation. Participants provided several examples of attempts to navigate the failures of the Lebanese authorities to provide identity documents and verification. For example, multiple participants said they held secondary passports, while one participant explained their choice to give birth to both their children in the United States, ``to give them another passport.'' Similar stories were iterated by other participants, pointing to how this particular form of infrastructuring was carried out by parents for the security of their children.

\section{Discussion}\label{sec:discussion}
In Lebanon, infrastructuring and security are tightly linked. This is partly a result of the simultaneous failures of distinct infrastructures, which necessitate infrastructuring work across multiple seams. Through such infrastructuring these seams are made visible, surfacing the security and insecurity that those living in Lebanon have to navigate on a daily basis. In this section we bring our findings into conversation with relevant CSCW and HCI infrastructuring literature as well as everyday security scholarship. We do so by conceptualising \emph{security patchworking}, developed and supported through our analysis. We use the conceptualisation of security patchworking to deepen our findings in Section~\ref{sec:dis-infrastructuring}, before setting out potential future directions for CSCW in designing for and with security patchworking in Section~\ref{sec:dis-future-directions}. 

\subsection{Security Patchworking: Layering Infrastructuring and Everyday Security in Lebanon}\label{sec:dis-infrastructuring}
The practices observed in our study are similar to those highlighted in existing research on infrastructuring (see Section~\ref{sec:rw-infrastructuring}), yet, they also differ in nuanced ways when brought into dialogue with everyday security. The infrastructuring work within HCI and CSCW research typically pertains to discrete infrastructures and processes undertaken by people to complete and/or work towards a goal affected by an infrastructure or wider service failing~\cite{CHI:GuiChe19,CHI:CVGMBBHP2019,joshi2021flaky, peer2022work} as well as to (re-)gain ontological security~\cite{semaan2019routine} and agency~\cite{holten2019assembling}. It is through observing the ways through which individuals~\cite{holten2019assembling}, groups~\cite{peer2022work,HCI:NtoKouVla21} and communities~\cite{joshi2021flaky,hussain2020infrastructuring} form varying assemblages in Lebanon that the failure of multiple infrastructures, ranging from the material to identity, become visible and as such seamful. In our research, we explore infrastructuring work that took place in and across multiple systems that overlap, connect and disconnect~\cite{vertesi2014seamful,CHI:SinJac17} rather than being tied to the failure of one system~\cite{joshi2021flaky}. Another central dimension of our work is our observation that the infrastructuring taking place in Lebanon was grounded in people's everyday experiences of insecurity, which led to routine practices to (re)establish security for themselves and others. 

Through this layering we conceptualise \emph{security patchworking} as a lens for analysing individual and collective security responses to debilitating infrastructures. Such responses comprise a series of patches for securing seamfully~\cite{CHI:ChaKum18}. Security patchworking thus orients itself towards meeting a specific goal intimately tied to the failing infrastructure (e.g., securing electricity) as well as working towards everyday security as populations create patches to simultaneously secure material, digital and identity needs. We propose security patchworking as a lens through which the interplay between multiple infrastructuring work and security are made visible: i) between security and insecurity within infrastructuring processes, ii) between ground-up and top-down infrastructuring for security, and iii) between infrastructuring identity and everyday security.

\subsubsection{Security and Insecurity of Infrastructuring}\label{sec:dis-sec-insec} Through the lens of security patchworking we observe that while infrastructuring contributes to everyday security through securing access to resources and services it also foregrounds forms of insecurity, demonstrating the (security) fragility of the patches made through infrastructuring. In our data, the reliance on generator cartels to secure electricity (see Section~\ref{sec:infrastructuring-electricity}) and on intermediaries to secure currency exchange (see Section~\ref{sec:infrastructuring-ponzi}) constitute forms of infrastructuring. However, while infrastructuring work may secure electricity in the short term, it generates increased financial insecurity as highlighted in our findings. This was also observed in how securing food vouchers was managed by voting for traditional political parties at the expense of securing the political imaginaries of the revolution (see Section~\ref{sec:infrastructuring-ponzi}). Our findings show that while such infrastructuring secures the immediate needs in Lebanon, they also contribute to a sense of long-term insecurity in people's everyday, which led some people in Lebanon to undertake further infrastructuring to secure foreign passports for their children. Such findings also draw parallels to~\cite{CSCW:TCJBGGAAM20}, where the authors found that while refugees worked around restrictions of food aid systems to secure non-food items, they were exposed to forms of insecurity created by these workarounds.

Security patchworking as a concept thus allows us to make sense of the infrastructuring work that is done to secure resources and services. As emphasised by one participant in Section~\ref{sec:infrastructure}, the demand for privacy wanes when access to food is scarce. This echoes findings in~\cite{COMPASS:SKABPCGD19} which highlight how the need to access aid overrides security and privacy concerns over the sharing of personal information and biometric data. Further, the failures of basic infrastructures in Lebanon deem privacy to be a luxury as also noted by HCI researchers working with and within marginalised communities~\cite{SOUPS:SCBANGMCC18,VasAndMar2018} and communities that employ intermediaries to access services and/or complete tasks~\cite{ICTD:GCMA15}. However, within those contexts, the individuals are often supported within trusted relationships that are rooted in familial relations and/or within intermediate services offered by banks~\cite{SIGCHI:SCTN10}, rather than unknown contacts established through an instant messaging service such as WhatsApp (as was experienced by the researcher when needed to exchange Lebanese currency in Beirut). Our findings also show how poor telecommunication infrastructures and the reduction in the availability of 3G mobile phone networks were responded to through infrastructuring that gave rise to insecurity (see Section~\ref{sec:spatial-security}). This in particular led to the most marginalised, such as refugees, having to rely on intermediaries for access with little agency over their own data. Echoing~\cite{NaaSim2021}, our findings also showed how restrictions and privatisation of public spaces in Lebanon led to infrastructuring across online and offline spaces. Participants reported that they resorted to online spaces for coming together and expressing their frustrations. However, this infrastructuring led to further insecurity through experiences of surveillance and persecution by the Lebanese authorities (see Section~\ref{sec:spatial-security}), leading to self-censorship similar to that reported in~\cite{HCI:JCJD21,CHI:OITD19}.

\subsubsection{Ground-Up, Top-down Security Patchworking}\label{sec:dis-GPTP} By understanding the infrastructuring work enacted by people in Lebanon to establish and maintain everyday security, our research also makes visible the mechanisms through which the Lebanese State secures itself. This brings understandings of the interplay between top-down~\cite{veeraraghavan2021cat} and ground-up patchworking~\cite{fox2023patchwork} into conversation with one another. The mode of governance maintained by the Lebanese political elite further results in the multiple infrastructural failures. The willful deterioration of infrastructures over recent years~\cite{WeiSow2020,RoaAls2021} and the Lebanese Government's refusal to reform these sectors has eroded essential services. This can be attributed to what Parreira~\cite{Parreira2020} refers to as the `Art of not Governing', where the Lebanese State's central power is strengthened not through authoritarian means from the centre, but by enabling and maintaining unregulated peripheries and processes (i.e., top-down patchworking~\cite{veeraraghavan2021cat}) that function in the seams of failing infrastructures. 

In our study, we adopted a ground-up examination of ongoing infrastructuring work around these failing infrastructures. However, layering this examination with everyday security makes visible the ways through which bricoleurs~\cite{peer2022work} strengthened the State's power by taking on the roles of unregulated peripheries and in turn contributing to experiences of insecurity (see Section~\ref{sec:dis-sec-insec}). Thus, we can observe the duality of bricoleurs as they are situated within both the ground-up patchworking for everyday security done by people in Lebanon as well as within the top-down patchworking in place to secure the State's power. This is exemplified by the generator cartels leveraging people's reliance on their energy supply to charge high monthly subscription rates for their services, while abusing clients and their competitors~\cite{Klinken2022}. This form of infrastructuring shifts the imperatives for securing essential services from the State to the peripheries, in an unregulated workaround that brings about new forms of insecurity. In contrast to `guerilla infrastructuring'~\cite{CHI:VCWO18}, peripheries in the context of Lebanon are not always necessarily negotiating with the centres of power (i.e., the Lebanese State). Rather, they are strengthening the centres of power by contributing to people's lived experience of insecurity. The same could be said about the unknown bricoleurs maintaining the platforms that people in Lebanon use to identify money exchange rates (see Section~\ref{sec:infrastructure}). Such platforms have taken on the economic and financial role of the State that previously set, pegged and communicated monetary exchange rates. However, the lack of transparency of how the rate is set contributes to rate fluctuation creating financial uncertainty and, in turn, experiences of insecurity~\cite{press:FP21}. In the same vein, the authors of~\cite{CSCW:ArmTalVla22} highlight how the infrastructuring work of the Lebanese diaspora supporting the 2019 revolution was pre-occupied with finding ways to work around the uncertainty of the currency. They describe how this pre-occupation came at the expense of the transnational public deliberating and negotiating towards a political agenda to counter the current political elite.  

\subsubsection{Infrastructuring Identity} Security patchworking deepens our understanding of the relation between infrastructuring, identity and security. In the context of Lebanon, everyday insecurity was grounded in long-term uncertainty that needed extensive infrastructuring work to ensure that basic security needs were met. This included the lack of State-governed security. As others have noted (see, e.g.,~\cite{CHI:LinNaaboy14,CHI:DasSem22,CHI:DosSem18}), the need to align with particular political views and/or identities to secure a livelihood often leads to different forms of identity work. In Lebanon, identity work also took the form of infrastructuring with people aiming to align with sectarian views -- and in turn the traditional political parties -- that were seen to provide some security, most often in a material sense. The re-positioning of identity and political priorities has been identified in other conflict settings. For example, Bj{\o}rn and Boulus-R{\o}dje~\cite{TOCHI:BjoBou18} found that Palestinian tech entrepreneurs were working within a cycle of continually having to reposition themselves as they manoeuvred infrastructural inaccessibility. As shown through our study, ontological security was pre-dominantly obtained through collective forms of identity work. For example, this was observed through assemblages such as the ecosystems or bubbles which Lebanese people, including the diaspora, built online to obtain and maintain their everyday security and sense of belonging. Here, the failure of the Lebanese State to authenticate an individual through an identification document -- verifying them as a \emph{real} person -- brought into conversation governance and ontological security, where the former cannot serve the latter. Put differently, Lebanese people were unable to establish what Steele~\cite{steele2005} refers to as ``security as being''.\\

\noindent In this section, we have brought our findings into conversation with infrastructuring and security scholarship. By conceptualising security patchworking we have made visible how infrastructuring work is practised for and with security within daily interactions and settings in Lebanon. 

\subsection{Future Directions for CSCW}\label{sec:dis-future-directions}
Here we concretise the implications of our research for CSCW, HCI and security researchers by suggesting future research directions that speak to the findings from our work.

\subsubsection{Infrastructuring through collective security} Coming together online and offline was a means of securing a collective sense of identity (see Section~\ref{sec:identity-security}) -- and, in turn, ontological security by establishing a shared identity (i.e, identity work). Indeed, infrastructuring for identity constituted infrastructuring work aimed at accessing public spaces to come together through protests as well as coming together online to grieve (see Section~\ref{sec:spatial-security}). Furthermore, the history -- and presence -- of sectarian divides had a direct impact on people's ontological security as they performed identity work to align with sectarian views that were seen to provide some form of security, most often in a material sense. It is within this context that we view infrastructuring for identity -- and in turn ontological security -- to be a collective endeavour that requires trust relations for this infrastructuring to be successful. Here, collectivity supports routine infrastructuring for ontological security, as presented by Semaan~\cite{semaan2019routine}, to be applied at scale and at a group rather than at an individual level. For example, our findings show how infrastructuring digital access -- integral to coming together online -- constituted processes and assemblages that relied on relations (e.g., Mukthars and neighbours) to secure electricity and Internet connectivity. Without such relations facilitating the infrastructuring work, these processes and assemblages would not yield the necessary or expected benefit for those undertaking the infrastructuring. Furthermore, each infrastructuring process relied on various bricoleurs with distinct political agendas~\cite{peer2022work}, thus also negotiating and navigating the intersections of top-down and ground-up infrastructuring work (see Section~\ref{sec:dis-GPTP}). Within this context of top-down and ground-up processes, CSCW researchers should consider drawing on adversarial design~\cite{disalvo2015adversarial}, where technologies are specifically designed to create spaces for confrontation and negotiation. In the context of our work, such spaces of conflict and negotiation -- and the technologies that mediate them -- would support the collective negotiation of everyday security and, in turn, shape collective security aspirations and mechanisms among Lebanese populations.

\subsubsection{Designing for assets and networks of support} As we outline in Section~\ref{sec:dis-infrastructuring}, the lens of security patchworking enables us to make visible the forms of insecurity that materialise through infrastructuring. Similar to~\cite{CHI:SinJac17}, our work shows how infrastructuring within seams create a process of inclusion and exclusion, i.e. those that are unable to navigate the seams find themselves actively excluded from the potential benefit of successful infrastructuring. Here, we can also invoke Dunne's~\cite{dunne2004we} notion of `security haves' and `security have-nots' to highlight the distinction between populations experiencing insecurity (the security have-nots) and those in a position of security (the security haves). In the context of Lebanon, the security have-nots are people who are unable to pay the generator cartels, for solar panels, Internet access or travel to give birth in another country; they found themselves -- in contrast to the the `security haves' -- unable to infrastructure towards everyday security. This requires us to consider how infrastructuring as a process may be designed for inclusivity in security terms. While security scholars typically refer to inclusivity in terms of designing for usability and accessibility, we suggest engaging in inclusive design that draws on socio-technical mechanisms of `networks of support'~\cite{coles2022protecting}. In the Lebanese context, this supports current practices such as the use of intermediaries to make current infrastructuring processes more inclusive. A design process that builds on existing support networks -- including processes and assemblages of people, objects and technologies -- while also mitigating the insecurities that may arise when relying on intermediaries, has the potential to foreground assets rather than barriers to everyday security. Leveraging assets within assets-based design approaches enables communities to continuously learn and reflect upon their assets and consider how they may mobilise them to attain their aspirations~\cite{CSCW:WGTD21}.

\subsubsection{Aspirational design} While existing CSCW and HCI work has shown how everyday security interplays with technological infrastructures such as food aid systems~\cite{CSCW:TCJBGGAAM20} and digitally enabled support networks~\cite{CHI:ColJenTal18}, our study also speaks to hopeful and aspirational design more broadly. Grounded in our findings, the infrastructuring work taking place in Lebanon should not only be understood as a means to meeting short-term needs -- but also as a process for people to work towards longer-term security goals and imaginaries, which were often suppressed due to everyday insecurity (see Section~\ref{sec:infrastructuring-ponzi} on the selling of electoral votes for survival rather than long-term political aspirations). Thus, we call on CSCW, HCI and security researchers and practitioners to consider drawing on aspirations-based design~\cite{Toyama2020}, where technological design is not solely oriented towards meeting the direct and immediate needs of people and communities but rather is ``persistent and aiming for something higher than one's current situation''~\cite{Toyama2020}. In other words, designing for the persistent desire of those minoritised and marginalised to achieve meaningful longer-term security as part of their daily lives. Thus, in many ways, our call follows in the footsteps of~\cite{khovanskaya2018designing} in which the authors initiate a design agenda ``against the status quo.'' Drawing on feminist critiques, for example, such an approach is unapologetic in its political goals towards more equitable and sustainable futures, where technologies work to reshape societal structures and power dynamics. Here, we view the need for the development of technologies that support the ways in which infrastructuring processes work towards aspirations of everyday security all the while securing access and services that are lacking due to infrastructural failures. By understanding our findings through lens of security patchworking we orient ourselves toward supporting distinct security imaginaries grounded in the lived experiences and aspirations of populations living in post-conflict contexts. This requires asking how current forms of infrastructuring and the technological tools they leverage may break the perpetual cycle of infrastructuring that results in new forms of insecurity. 

\subsubsection{Transparency, verification and authentication} Our findings show that people in Lebanon often relied on unverified platforms for money exchange rates that lacked transparency. Similarly, the informal sector of securing electricity from what participants termed `generator cartels' was unregulated, leaving the market open to profiteering without transparent reporting on market prices and cost justifications. We propose collaborations between CSCW and security researchers to work with communities to adopt design approaches that leverage assets for infrastructuring work for transparency, verification and authentication. Additionally, enhancing transparency, verification and authentication holds the potential to disrupt the ways in which unregulated peripheries strengthen the State. In other words, potentially reversing the power relationship between competing top-down and ground-up infrastructuring. Additionally, CSCW and security researcher~\cite{sultana2021dissemination,kaiser2021adapting, kirchner2020countering,haque2020combating} have drawn up guidelines and recommendations that support users in the verification of online information and/or the design of technologies and interfaces that help them do so. In the context of Lebanon, the unverified and non-transparent currency exchange platforms are enmeshed within the infrastructuring processes that contribute to everyday security and, by extension, insecurity. Therefore, we call for transparency, verification and authentication mechanisms that provide not only material re-assurances but also emotional ones through security infrastructures of care~\cite{tseng2022care}. 

\section{(Dis)conclusion}\label{sec:conclusion}
Our research points to how people in Lebanon are continuously required to infrastructure towards securing the material, digital and identity in response to failing infrastructures. We show how these infrastructuring processes contribute to everyday and ontological security. We contribute \emph{security patchworking} as a lens for layering infrastructuring with everyday security and in doing so making visible the interplay between security and insecurity within infrastructuring in Lebanon, ground-up and top-down infrastructuring as well as the multiple facets of infrastructuring identity and in-turn ontological security. We then discuss future work that acts on what is made visible through security patchworking. In the process of writing this paper, we ourselves were continuously re-writing elements of it to secure the anonymisation and safety of participants, to secure the ability to go back home for those of us who are Lebanese and the trustful and valued relationships that the researcher built while in Beirut. As such we have been engaging in our own acts of everyday security as we navigate the insecurities experienced in Lebanon. The infrastructuring in Lebanon has not concluded and as such we would be remiss to offer a conclusion ourselves, fore underneath the frustrations voiced by participants and felt through multiple encounters, a sense of resistance also characterised people's responses to infrastructural failures and a lack of security.

\subsection{Limitations}\label{sec:limitations}
A number of limitations should be taken into account when interpreting our findings. First, the fieldwork took place over a two-week period in Beirut and thus has the potential to be expanded both in time, scope and geography. Second, interviews were carried out with people who had a leading voice in political debates in Lebanon and hence the work can be expanded to work with different societal groups who are not `experts' in the field of digital rights and democracy. Third, the level of participation has limitations such as language barriers~\cite{lipson1989}. While one of the authors speaks Arabic, the fieldworker does not, although this -- along with French and English -- is the dominant language in Lebanon. Finally, as noted in Section~\ref{sec:methodology}, there is an inherent bias in research relying on interviews and focusing on security and/or politically charged questions, given that participants self-select to take part. It could be that some participants decided against participation as a result. However, we tried to overcome this limitation by immersing ourselves in the setting through an ethnographically informed approach.

\section*{Acknowledgements}
This work would not exist without the many contributions from people, both in and outside Lebanon, who so generously gave of their time to speak to us and who have remained engaged with the research throughout. We thank the anonymous reviewers for their constructive feedback. The research of McClearn was supported by the EPSRC as part of the Centre for Doctoral Training in Cyber Security for the Everyday at Royal Holloway, University of London (EP/S021817/1).

\bibliographystyle{plain}
\bibliography{local}

\end{document}